\documentclass[letterpaper, 10 pt, conference]{ieeeconf}
\IEEEoverridecommandlockouts
\usepackage{ifpdf}
\usepackage[frozencache=true,cachedir=minted-cache]{minted} 

\usepackage{bm}
\usepackage{amsmath,amssymb,amsfonts}
\usepackage{algorithmic}
\usepackage{graphicx}
\usepackage{textcomp}
\usepackage{xcolor}
\usepackage{gensymb}
\usepackage{comment}
\usepackage{tikz}
\usepackage[utf8]{inputenc}
\usepackage{pgfplots} 
\usepackage{pgfgantt}
\usepackage{pdflscape}
\usepackage{csquotes}
\usepackage{adjustbox}
\usepackage{booktabs}
\usepackage{multirow}

\pgfplotsset{compat=newest} 
\pgfplotsset{plot coordinates/math parser=false}
\def\BibTeX{\scriptsize{\rm B\kern-.05em{\sc i\kern-.025em b}\kern-.08em
    T\kern-.1667em\lower.7ex\hbox{E}\kern-.125emX}}
\newcommand{\figref}[1]{Fig.~\ref{#1}}

\title{\LARGE \bf
Physics-Constrained Taylor Neural Networks\\for Learning and Control of Dynamical Systems}
\author{Nam T. Nguyen$^{1}$ and Juan C. Tique$^{1}$
\thanks{This material is based upon work supported by the  SGS 2024 program at Northern Arizona University for Nam Nguyen.}
\thanks{Corresponding author: {\tt\small ntn44@nau.edu} (Nam T. Nguyen)}
\thanks{$^{1}$School of Informatics, Computing, and Cyber Systems, Northern Arizona University, Flagstaff, AZ, USA.}
}
\begin{document}

\maketitle
\thispagestyle{empty}
\pagestyle{empty}
\begin{abstract}
Data-driven approaches are increasingly popular for identifying dynamical systems due to improved accuracy and availability of sensor data. However, relying solely on data for identification does not guarantee that the identified systems will maintain their physical properties or that the predicted models will generalize well. In this paper, we propose a novel method for system identification by integrating a neural network as the first-order derivative of a Taylor series expansion instead of learning a dynamical function directly. This approach, called Monotonic Taylor Neural Networks (MTNN), aims to ensure monotonic properties of dynamical systems by constraining the conditions for the output of the neural networks model to be either always non-positive or non-negative. These conditions are constructed in two ways: by designing a new neural network architecture or by regularizing the loss function for training. The proposed method demonstrates better performance compared to methods without constraints on the monotonic properties of the systems when tested with experimental data from two real-world systems, including HVAC and TCLab. Furthermore, MTNN shows good performance in an actual control application when using a model predictive controller for a nonlinear MIMO system, illustrating the practical applications of this method.
\end{abstract}

\section{Introduction}

Machine learning (ML) approaches are becoming increasingly popular for identification tasks in non-linear dynamical systems due to advancements in sensor technology and the development of ML algorithms for these systems \cite{bruntonDataDrivenScienceEngineering2022}. Nevertheless, relying solely on data can lead to issues such as a loss of generalization when predicting system behavior \cite{karniadakisPhysicsinformedMachineLearning2021a}. Additionally, without constraints of physical properties, black-box models might predict signals that fall outside safe ranges, potentially causing system failures caused by prediction errors. To address these challenges, the field of physics-informed machine learning has emerged, combining model-based and data-driven approaches to enhance prediction reliability and maintain system safety \cite{karniadakisPhysicsinformedMachineLearning2021a}. 

Leveraging these advantages, this paper explores the integration of monotonic and convex properties into machine learning models, which have significant practical implications, particularly in the field of dynamical systems. A common approach for enforcing monotonicity in neural networks (NNs) is to constrain the weights to be either non-negative or non-positive \cite{archerApplicationBackPropagation1993}. While this ensures monotonicity, it often leads to suboptimal performance by restricting the network to be perpetually convex. Another established architecture designed to impose monotonicity is the Min-Max network \cite{danielsMonotonePartiallyMonotone2010a}, which adheres to the universal approximation theorem. However, due to the extreme nonlinearity of the Min and Max functions, this method frequently yields uncertain model parameters when trained on small datasets. The Deep Lattice Network (DLN) introduces a distinct class of functions—ensembles of lattices \cite{youDeepLatticeNetworks}—as differentiable elements within the NN architecture. Nevertheless, DLNs typically require a large number of parameters to achieve satisfactory results, thereby necessitating a considerable amount of training data. The monotonic neural ODE method \cite{monODE} has demonstrated promising results in function approximation, though its application has been proposed to time-series data without exogenous inputs, and no control applications have yet been explored using this approach.

This paper presents a novel method that imposes monotonicity constraints and potentially guarantees convexity constraints on NN models to improve their performance in complex systems. The approach leverages Taylor series expansion to approximate functions from sequential data, focusing on learning first-order derivatives rather than directly modeling the entire function through the NN model. This enables the network to inherently preserve monotonicity by constraining either its architecture or outputs without requiring additional derivative calculations. In contrast, directly modeling the dynamic function requires computing and constraining the neural network’s derivatives—a process that is highly dependent on data quality for accurate differentiation. By employing the Taylor series, our method is particularly well-suited for systems with external inputs, multiple states, and multiple outputs, as this technique efficiently approximates multivariable functions. 

To enforce monotonicity, we constrain the output of the NN model to be non-positive or non-negative, depending on whether the relationship between inputs and outputs is decreasing or increasing. There are two approaches to ensure monotonic conditions: inductive bias and learning bias. In the inductive bias approach, the NN model has built-in constraints that ensure its outputs are always non-positive or non-negative in response to monotonic inputs. In the learning bias approach, these constraints are incorporated as regularization terms in the loss function during training. We refer to this technique as Monotonic Taylor Neural Networks (MTNN). Otherwise, convexity can also be guaranteed by adding regularization terms that penalize the model for producing outputs that violate the desired convex properties. However, constraining convexity requires differentiating the neural network once, which is suitable for the second-order approximation of the Taylor series.  

In order to demonstrate the capabilities of our proposed method, we apply it to experimental data from actual Heating, Ventilation, and Air Conditioning (HVAC) systems for identification tasks. The results show that MTNN produces more robust and accurate outcomes than other methods, including non-constrained Taylor neural networks, vanilla neural networks, and the Min-Max model, even when trained on small datasets. Furthermore, we implemented the method to design a model predictive controller (MPC) for a multiple-input multiple-output (MIMO) system, which was then tested in the real-world Temperature Control Laboratory (TCLab). The controller also demonstrated effective performance, with system outputs successfully tracking the reference setpoints, even with huge differences from initial states to references.

The remainder of this paper is structured as follows: Section I outlines the application of Taylor approximations in dynamical systems. Section III details the construction of monotonic properties within the neural network architecture and the corresponding loss function. Section IV describes the setup of the model predictive controller for the MTNN model. Section V presents the results for the identification of the HVAC system and the results of a practical control application in a MIMO system. Finally, the paper concludes with a summary of the findings of this work.

\section{Learning Taylor Neural Networks for Dynamical Systems}
In a dynamical system, the characteristics of the system are presented by the ordinary differentiable equation (ODE):
\begin{align}
    \frac{d\mathbf{x}}{dt} & = \text{ODE}(\mathbf{x}, \mathbf{u}) \label{ODE}
\end{align}
where $\mathbf{x} = [x_1, x_2, ..., x_{Nx}]^T$ is a vector of the states and $\mathbf{u}  = [u_1, u_2, ..., u_{Nu}]^T$ is a vector of the exogenous inputs of the system. This paper identifies the dynamical system by considering its discrete-time inputs and outputs. The discrete version of the model in \eqref{ODE} is expressed as:
\begin{align}
    \mathbf{x}^{k+1} = f(\mathbf{x}^{k}, \mathbf{u}^{k})\label{disODE}
\end{align}
where $k$ represents time step $k$. Here, we also assume that all states in vector $\mathbf{x}$ are measurable or observable.

We approximate \eqref{disODE} by using Taylor-series expansion as the second order expansion about the expansion point $[\mathbf{x}^{k*}; \mathbf{u}^{k*}]$ as in equation \eqref{Taylor_2nd_modified}.
\begin{align}
    \mathbf{x}^{k+1} &\approx f(\mathbf{x}^{k*}, \mathbf{u}^{k*}) + 
    \mathcal{J}^{*} 
    \begin{bmatrix}
    \Delta \mathbf{x}^{k*} \\
    \Delta \mathbf{u}^{k*}
    \end{bmatrix} 
    + \frac{1}{2} \begin{bmatrix} 
    \Delta \mathbf{x}^{k*} \\
    \Delta \mathbf{u}^{k*}
    \end{bmatrix}
    \mathcal{H}^{*} 
    \begin{bmatrix} 
    \Delta \mathbf{x}^{k*} \\
    \Delta \mathbf{u}^{k*}
    \end{bmatrix}^T  \label{Taylor_2nd_modified}
\end{align}
where $\mathcal{J}^{*}$ and $\mathcal{H}^{*}$ are the Jacobian matrix and Hessian matrix of function $f$ at $[\mathbf{x}^{k*}; \mathbf{u}^{k*}]$; $[\Delta \mathbf{x}^{k*}; \Delta \mathbf{u}^{k*}]^T = [\mathbf{x}^{k} - \mathbf{x}^{k*}; \mathbf{u}^{k} - \mathbf{u}^{k*}]^T$. With the optimal expansion points $[\mathbf{x}^{k*}; \mathbf{u}^{k*}]$, where it helps $f([\mathbf{x}^{k*}; \mathbf{u}^{k*}])$ converge quickly to $f([\mathbf{x}^{k}; \mathbf{u}^{k}])$ (or $\mathbf{x}^{k+1}$), the function can be approximated accurately by the first-order approximation of Taylor series. 
\begin{figure}[b]
    \centering
    \includegraphics[trim=0.4cm 0.cm 0.cm 0.cm, clip]{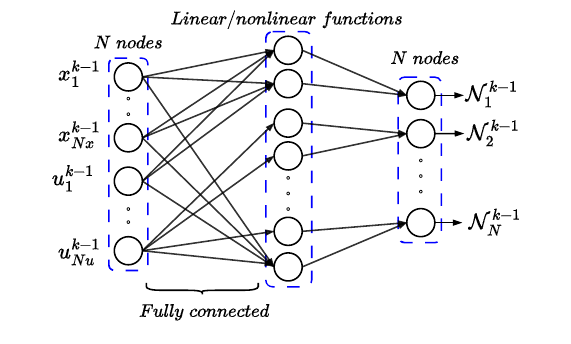}
    \caption{Neural network of first-order derivative}
    \label{fig:NN-deri}
\end{figure}

If the time-step measurement of a dynamical system is small enough, or the sampling rate is sufficiently high, the previous inputs can be considered as the optimal expansion points for the Taylor series \cite{hanselmannLearningFunctionsTheir1999}. Additionally, higher-order Taylor series approximations provide greater accuracy. Hence, using previous points, Taylor expansion remains efficient if the series orders are high enough for fast dynamic systems. For this reason, in this paper, we select the previous data as the expansion points $[\mathbf{x}^{k-1}; \mathbf{u}^{k-1}]$ for our Taylor neural networks model. To write algorithms easily, we set $\mathbf{z}^{k-1} = [\mathbf{x}^{k-1}; \mathbf{u}^{k-1}]$; $\mathbf{z}^{k-1} = [z_1^{k-1},..., z_N^{k-1}]$. Next, we can utilize the second-order Taylor series to calculate $\mathbf{x}^{k+1}$ with the expansion point is at the previous step $k-1$:
\begin{align}
    \mathbf{x}^{k+1} &\approx \mathbf{x}^{k} + 
    \mathcal{J}(\mathbf{z}^{k-1}) 
    \Delta \mathbf{z}^{k}
    + \frac{1}{2} 
    \Delta \mathbf{z}^{k}
    \mathcal{H}(\mathbf{z}^{k-1})(\Delta \mathbf{z}^{k})^T
  \label{Taylor_2nd_previous}
\end{align}
where $\Delta \mathbf{z}^{k} =[\mathbf{z}^{k} - \mathbf{z}^{k-1}] = [\mathbf{x}^{k} - \mathbf{x}^{k-1}; \mathbf{u}^{k} - \mathbf{u}^{k-1}]^T$.

Instead of determining the function by using the NN models directly, this paper adopts an NN model to approximate the derivative of the dynamical model. We will model the Jacobian matrix as a neural network with the input $\mathbf{z}^{k-1}$.
\figref{fig:NN-deri} shows how the neural network is constructed to replace the Jacobian matrix. We have an MTTN model to predict the next states of the system:
\begin{align}
    \widehat{\mathbf{x}}^{k+1}& =\text{MTNN}(\mathbf{x}^{k}, \mathbf{u}^{k}, \mathbf{x}^{k-1}, \mathbf{u}^{k-1}) \notag \\
    &  = \text{MTNN}(\mathbf{z}^{k},\mathbf{z}^{k-1})\label{Taylor_NN}\\
    &  =  \mathbf{x}^{k} + \bm{\mathcal{N}}(\mathbf{z}^{k-1})
    \Delta \mathbf{z}^{k}
    + \frac{1}{2} 
    \Delta \mathbf{z}^{k} \frac{\partial \bm{\mathcal{N}}}{\partial \mathbf{z}}\bigg|_{\mathbf{z}^{k-1}} 
    (\Delta \mathbf{z}^{k})^T \notag\\
    \bm{\mathcal{N}}(\mathbf{z}^{k-1})& = \begin{bmatrix}
    \mathcal{N}(\mathbf{z}^{k-1}) \big|_{\theta_{1}} \\ \mathcal{N}(\mathbf{z}^{k-1}) \big|_{\theta_{2}}  \\
    \cdots \\
    \mathcal{N}(\mathbf{z}^{k-1})\big|_{\theta_{Nx}} 
    \end{bmatrix} (\bm{\mathcal{N}}(\mathbf{z}^{k-1}) \in \mathbb{R}^{N_x*N}) \label{jacobian} 
\end{align}
where each neural network with its parameters $\theta_j (j = 1,..., N_x)$, $\mathcal{N}|_{\theta_j}$ in \eqref{jacobian} has $N$ inputs and $N$ outputs ($N = N_x + N_u$). With MIMO systems, MTNN will be implemented for each output. For example, if the system has $N_o$ outputs, we will use $N_o$ models as in \figref{fig:NN-deri}. The Hessian matrix, denoted by $\partial \bm{\mathcal{N}} / \partial \mathbf{z}$, is defined as in equation \eqref{Hessian}.
\begin{align}
    \frac{\partial \bm{\mathcal{N}}}{\partial \mathbf{z}} & = \left[ \frac{\partial \bm{\mathcal{N}}}{\partial z_1} \;\frac{\partial \bm{\mathcal{N}}}{\partial z_2} \; \cdots \; \frac{\partial \bm{\mathcal{N}}}{\partial z_N} \right] \label{Hessian} \\
    \frac{\partial \bm{\mathcal{N}}}{\partial z_i} &  =\left[
    \frac{\partial \mathcal{N}}{\partial z_i} \Big|_{\theta_1};\;
    \frac{\partial \mathcal{N}}{\partial z_i} \Big|_{\theta_2};\;
    \cdots;\;
    \frac{\partial \mathcal{N}}{\partial z_i} \Big|_{\theta_{N_x}} 
\right]
\end{align}

When using higher-order approximations of Taylor neural networks (higher than first-order), we need to calculate the derivatives of the neural network with respect to its inputs. Currently, PyTorch \cite{paszkeAutomaticDifferentiationPyTorch} and TensorFlow \cite{pangDeepLearningTensorFlow2020} libraries support calculating neural network gradients with high accuracy. Hence, this paper uses these automatic differentiation libraries to compute higher-order derivatives of NN models.
It is worth mentioning that when using higher-order approximations of Taylor neural networks, activation functions that are infinitely differentiable, such as $tanh()$ and $sigmoid()$, are preferred. On the other hand, for systems where dynamic functions are of degree one or two, it is more efficient to use linear functions in place of nonlinear activation functions within the neural network architecture. This approach leads to faster training and reduces the amount of training data.

\textbf{Learning MSTNN model:} To train MSTNN model, this paper utilizes mean square errors (MSE) as the primary loss function, which is defined in \eqref{mse}:
\begin{align}
    \mathcal{L}_{MSE} &  =\frac{1}{N_{train}}\sum_{k = 0}^{N_{train}}(\mathbf{x}^{k+1} - \widehat{\mathbf{x}}^{k+1})^2 \label{mse} \\
    \widehat{\mathbf{x}}^{k+1} & = \text{MTNN}(\mathbf{x}^{k}, \mathbf{u}^{k}, \mathbf{x}^{k-1}, \mathbf{u}^{k-1}) 
\end{align}
where \(N_{train}\) is the total number of training data points, and \(\mathbf{x}^{k+1}\) and \(\widehat{\mathbf{x}}^{k+1}\) are the observed outputs and predicted outputs, respectively.
\section{Monotonicity and Convexity Constraints in Neural Networks}
This section will present how we constrain the monotonicity of the Taylor neural networks model in two ways: create a new NN architecture and regularize the loss function. Besides dealing with convex systems, convexity is also guaranteed as a regularization in the loss function.
\subsection{Monotonicity}
To enforce monotonicity, we constrain the outputs $\mathcal{N}_i([\mathbf{x}^{k}; \mathbf{u}^{k}]) , (i = 1,\dots,N)$ to be either non-positive or non-negative, depending on whether the partial monotonic relationship between the inputs and the output $f^{k}$ is decreasing or increasing, respectively. This section introduces two methods for incorporating monotonicity into the dynamic functions: (1) by designing the neural network architecture directly, and (2) by adding monotonicity constraints as regularization terms in the loss function.

\subsubsection{Constraints on Neural Network Architecture}
\begin{figure}[t]
    \centering
    \includegraphics[scale = 1.05, trim=0.2cm 0.2cm 0.3cm 0, clip]{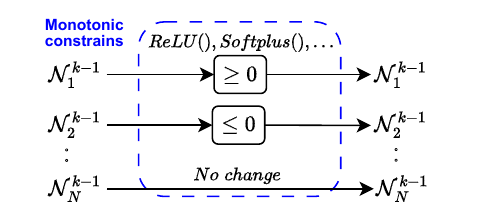}
    \caption{Monotonic property constrain}
    \label{fig:mono}
\end{figure}

For increasing monotonicity, when inputs increase, the corresponding derivatives of the function with respect to those inputs must be non-negative (with zero derivatives for constant inputs). Conversely, for decreasing monotonicity, the corresponding derivatives must be non-positive. As illustrated in \figref{fig:mono}, we can enforce these conditions by constraining the derivatives to be greater than or equal to zero for increasing monotonicity. For instance, if $\mathcal{N}({z_1^{k-1}})$, representing input $z_1$, exhibits increasing monotonicity, while $\mathcal{N}({z_2^{k-1}})$, representing input $z_2$, exhibits decreasing monotonicity, the respective constraints will be applied accordingly. For inputs where the monotonic behavior is unknown, no constraints are enforced, which we refer to as ''no change'' line.

To implement the constraint $\mathcal{N}({z_1^{k-1}}) \geq 0$, functions such as \texttt{\texttt{\texttt{ReLU()}}} can be used to ensure the output $\mathcal{N}({z_1^{k-1}})$ is always non-negative. Besides, other activation functions, like exponential or softplus functions, can also be employed to enforce this monotonicity condition. Similarly, to enforce decreasing monotonicity, we can constrain $\mathcal{N}({z_2^{k-1}})$ by applying $\mathcal{N}({z_2^{k-1}}) := -\text{ReLU}(\mathcal{N}({z_2^{k-1}})$, ensuring that this value always remains non-positive. Once these constraints are integrated into the outputs of \figref{fig:NN-deri}, the resulting neural network architecture will be used in \eqref{Taylor_NN} to predict the next states of the dynamical system.

\subsubsection{Constraints on Loss Function} Regularization will be added with the MSE loss, which imposes conditions on neural network outputs, either positive or negative, based on the monotonic relationship between inputs and their derivatives to guarantee monotonicity. The monotonic loss function is illustrated in \eqref{mon_mse}:
\begin{align}
    \mathcal{L}_{mono} &= \mathcal{L}_{MSE} + \sum_{j=1}^{N_{po}} \lambda_j \mathcal{M}(- \mathcal{N}_j) + \sum_{j=1}^{N_{ne}} \lambda_j \mathcal{M}(\mathcal{N}_j) \label{mon_mse}
\end{align}
where \(N_{po}\) and \(N_{ne}\) represent the number of positive derivatives and negative derivatives, respectively (or the number of positive outputs and negative outputs in \(\mathcal{N}\)). The regularization parameter is denoted as \(\lambda_j\) in the last two terms. In \eqref{mon_mse}, the \(\mathcal{M}()\) functions are implemented to ensure the monotonic properties. For example, for positive derivatives, we aim to force all negative outputs of \(\mathcal{N}\) to reach zero. Conversely, for constraining negative derivatives, we seek to push all positive outputs of \(\mathcal{N}\) toward zero. We can select the function \(\mathcal{M}()\) as a \texttt{ReLU()} function.

\subsection{Potential in Convex Constraints}

In addition to monotonicity, convexity is another important property in dynamical systems. For a function to be convex, its Hessian matrix must be positive semi-definite. Conversely, if the function is concave, the Hessian matrix will be negative semi-definite. Based on these characteristics, we can ensure the convexity and concavity of a function when training our MTSNN by using the following loss function.
\begin{align}
    \mathcal{L}_{conv} & = \mathcal{L}_{MSE} + \gamma\mathcal{C}\left(-\text{det}\left({\frac{\partial \bm{\mathcal{N}}}{\partial \mathbf{z}}}\right)\right) \label{con_loss}
\end{align}
where \(\det(\cdot)\) denotes the determinant of a matrix. \(\mathcal{C}()\) in \eqref{con_loss} operates similarly to \(\mathcal{M}()\) in \eqref{mon_mse}, that we can use $\texttt{\texttt{ReLU()}}$ function. Here, \(\gamma\) represents the regularization parameter that controls the weight of the convexity or concavity constraints. Also, it is important to note that the neural network's differentiation is computed with respect to the inputs. Therefore, the quality of the measured data must be high to ensure the accuracy of derivative calculations.
\section{Model Predictive Control for MTNN Model}
\begin{figure}[b]
    \centering
    \includegraphics[scale=0.93, trim=0.45cm 0.35cm 0.75cm 0.1cm, clip]{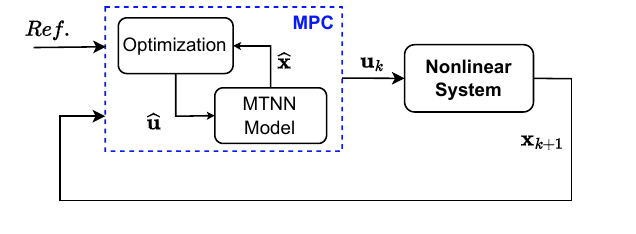}
    \caption{MPC controller for MTNN model}
    \label{fig:MPC}
\end{figure}
In order to demonstrate the accuracy of the MTNN model, this paper creates a controller by using the proposed model and implementing it in a real system (real temperature control laboratory \cite{rossiterEvaluationDemonstrationTake2019}). Due to the aspect of designing controllers for nonlinear systems, we can typically linearize systems around operating points. After linearization, it becomes easier to apply linear controllers such as PID \cite{nguyenEnhancedFuzzyMFCbasedTraction2023}. However, linearization in the context of machine learning models can be a challenging step and lacks truthfulness. Therefore, using nonlinear controllers can be more convenient and efficient in our case. 
In \cite{yangModelPredictiveControl2020}, the authors demonstrated that the MPC technique is a powerful controller for handling machine learning models. MPC consists of two main components: optimization and the predictive model, as shown in \figref{fig:MPC}. The optimization part determines the optimal control input by minimizing a cost function subject to constraints, as described in the equations (\ref{mpc_cost_function}-\ref{condition}).  
\begin{align}
    J =  & \sum_{k=0}^{N_{h}-1}  \left[ \| \mathbf{x}^{k} - \mathbf{x}_{ref} \|_{Q}^2 + \| \mathbf{u}^{k} \|_{R}^2 \right] + \| \mathbf{x}^{N_h} - \mathbf{x}_{ref}\|_{P}^2 \label{mpc_cost_function} \\
    \text{s.t.} \hspace{0.1 cm}  & \mathbf{x}^{k+1} =\text{MTNN}(\mathbf{x}^{k}, \mathbf{u}^{k}, \mathbf{x}^{k-1}, \mathbf{u}^{k-1})
    \\
    & \mathbf{x}_{\text{min}} \leq \mathbf{x}^{k} \leq \mathbf{x}_{\text{max}}, \hspace{0.2cm} \text{for } k = 0, \ldots, N_h
    \\
    & \mathbf{x}^0 = X_0 \quad \text{(this value is from historical data)} \\
    & \mathbf{u}_{\text{min}} \leq  \mathbf{u}^{k} \leq \mathbf{u}_{\text{max}} \hspace{0.2cm} \text{for } k = 0, \ldots, N_h-1\label{condition}
\end{align}
where $\mathbf{x}_{\text{ref}}$ denotes the reference state, while $\mathbf{u}_{\text{min}}$ and $\mathbf{u}_{\text{max}}$ represent the minimum and maximum limits of the control inputs $\mathbf{u}^{k}$, respectively. Similarly, $\mathbf{x}_{\text{min}}$ and $\mathbf{x}_{\text{max}}$ define the state constraints for $\mathbf{x}$. It is important to emphasize that in MPC, the accuracy of the predicted model is critical. Since the model must predict multiple steps over the horizon $N$, any prediction errors can accumulate over time, potentially degrading control performance. The cost function $J$ is formulated over an $N_h$-step horizon and is minimized to find the optimal vector of inputs for the entire horizon. This optimal input vector's initial input (time step $k = 0$ of vector $\mathbf{u}^k$) is then applied to the nonlinear system. In the next section, we will show the results of system identification and create an MPC model when using MTNN.

\begin{table*}[t]
\centering
\caption{$R^2$ and $RMSE$ results with five multi-step predictions in HVAC scenario.}
\label{table}
\scalebox{0.98}{ 
\setlength{\tabcolsep}{5pt} 
\renewcommand{\arraystretch}{1.01} 
\begin{adjustbox}{width=\textwidth}
\begin{tabular}{|c|c|c|c|c|c|c|c|c|}
\hline
\multicolumn{9}{|c|}{\vspace{-8.7 pt}} \\
\textbf{Step} & {\textbf{Baseline}} & {\textbf{Min-max}} & {\textbf{1st Taylor}} & {\textbf{2nd Taylor}} & {\textbf{1st Mono Taylor}} & {\textbf{2nd Mono Taylor}} & {\textbf{1st Soft Taylor}} & {\textbf{2nd Soft Taylor}} \\
\multicolumn{9}{|c|}{\vspace{-8.7pt}} \\
\hline
\multicolumn{9}{|c|}{\vspace{-8.7pt}} \\
\multicolumn{9}{|c|}{\textbf{R$\mathbf{^2}$ score}} \\
\multicolumn{9}{|c|}{\vspace{-8.7pt}} \\
\hline
1 & 0.9303 & 0.9426 & 0.9460 & 0.9620 & 0.9765 & \textbf{0.9765} & \textbf{0.9806} & 0.9801 \\ \hline \multicolumn{9}{|c|}{\vspace{-8.7pt}} \\
2 & 0.7940 & 0.8330 & 0.9179 & 0.9343 & 0.9532 & \textbf{0.9532} & \textbf{0.9577} & 0.9569 \\ \hline \multicolumn{9}{|c|}{\vspace{-8.7pt}} \\
3 & 0.6270 & 0.6945 & 0.6972 & 0.8710 & 0.9433 & \textbf{0.9435} & \textbf{0.9452} & 0.9437 \\ \hline \multicolumn{9}{|c|}{\vspace{-8.7pt}} \\
4 & 0.4507 & 0.5432 & 0.5010 & 0.8175 & 0.9098 & \textbf{0.9100} & \textbf{0.9174} & 0.9162 \\ \hline \multicolumn{9}{|c|}{\vspace{-8.7pt}} \\
5 & 0.2858 & 0.3992 & 0.3077 & 0.7744 & 0.9073 & \textbf{0.9077} & \textbf{0.9030} & 0.9034 \\
\hline
\multicolumn{9}{|c|}{\vspace{-8.7pt}} \\
\multicolumn{9}{|c|}{\textbf{RMSE ($^{\circ}$F)}} \\
\multicolumn{9}{|c|}{\vspace{-8.7pt}} \\
\hline

1 & 0.2403 & 0.2182 & 0.2116 & 0.1774 & 0.1397 & \textbf{0.1395} & \textbf{0.1267} & 0.1267 \\ \hline \multicolumn{9}{|c|}{\vspace{-8.7pt}} \\
2 & 0.4081 & 0.3674 & 0.2575 & 0.2305 & 0.1946 & \textbf{0.1945} & \textbf{0.1850} & 0.1850 \\ \hline \multicolumn{9}{|c|}{\vspace{-8.7pt}} \\ 
3 & 0.5439 & 0.4923 & 0.4900 & 0.3199 & 0.2120 & \textbf{0.2116} & \textbf{0.2085} & 0.2085 \\ \hline \multicolumn{9}{|c|}{\vspace{-8.7pt}} \\
4 & 0.6538 & 0.5962 & 0.6231 & 0.3769 & 0.2649 & \textbf{0.2646} & \textbf{0.2536} & 0.2536 \\ \hline \multicolumn{9}{|c|}{\vspace{-8.7pt}} \\
5 & 0.7388 & 0.6777 & 0.7274 & 0.4153 & 0.2662 & \textbf{0.2656} & \textbf{0.2723} & 0.2723 \\
\hline
\end{tabular}
\end{adjustbox}
}
\end{table*}
\section{Experimental results}
\subsection{System Identification of HVAC systems} 
\textbf{HVAC system}:
This paper focuses on the control of room temperature in HVAC systems, where the temperature is influenced by two primary inputs: the supply discharge temperature \(T_s\) and the mass flow rate \(\dot{m}\) from the variable air volume (VAV) system. In the discrete-time model, the next room temperature \(T^{k+1}\) is computed as a function of three inputs: the current room temperature \(T^{k}\), the current supply discharge temperature \(T_{s}^{k}\), and the current mass flow rate \(\dot{m}^{k}\). Notably, the relationship between \(T_{s}^{k}\) and \(T^{k+1}\) exhibits partial monotonicity, with the temperature increasing as a function of \(T_{s}^{k}\). Equation \eqref{HVAC} presents the relationship between them.
\begin{align}
    T^{k+1} = f_{\text{HVAC}}\left(T^{k}; T_{s}^{k}; \dot{m}^{k}\right) \label{HVAC}
\end{align}

The aforementioned model is considered as a multiple-input, single-output (MISO) system. To illustrate the identification results, actual data from the HVAC system at the School of Informatics, Computing, and Cyber Systems building at Northern Arizona University were utilized for training and testing the machine learning models. The data were collected during the cooling season in April 2021. The room under observation is centrally located within the building and serves as a large classroom with a seating capacity of eighty people. However, there are no students when collecting data.

The entire dataset, collected at 5-minute intervals, includes $\{\hspace{0.1cm} T^{k+1},\hspace{0.1cm} [T^{k}, \dot{m}^{k}, T_{s}^{k}]\}$. The experiments use $N_{train} = 180$ initial values from the dataset, where the temperature range in the training set is between $68^\circ F$ and $73^\circ F$. The testing data consists of $N_{test} = 100$ values, where the temperature range in the testing set is between $73^\circ F$ and $76^\circ F$. This scenario is particularly challenging due to the large step size and the different ranges between the training and testing sets. Therefore, it effectively demonstrates the generalization potential of the methods.

\textbf{Testing Metrics:}
Two conventional performance metrics are employed to evaluate the effectiveness of the proposed methodologies in comparison to existing approaches: the $R$ squared score ($R^2$) and the root mean square error ($RMSE$).

Within the model predictive control (MPC) framework, it is essential not only to compute accurately the next state of the system but also to forecast multiple subsequent states (up to $i$ steps ahead). This multi-step prediction is necessary for determining optimal control actions over a receding horizon in MPC. To validate the robustness of the proposed model, we perform predictions of the following $i$ state values, as represented in \eqref{next_hvac}.
\begin{align}
    \widehat{T}^{k+1+i} & = f_{\text{HVAC}}(\widehat{T}^{k+i}; T_s^{k+i}; \dot{m}^{k+i}) \label{next_hvac}
\end{align}
where $\widehat{T}^{k+i}$ denotes the predicted temperature at time step $k+i$ from the neural network model, and the corresponding variables, including $T_s^{k+i}$ and $\dot{m}^{k+i}$ refer to the actual measured data collected from the physical system.

\textbf{Results:}
This section compares the accuracy of eight approaches in terms of their ability to identify patterns. \texttt{"Baseline"} represents a standard artificial neural network that approximates the model directly from inputs to outputs. \texttt{"Min-max"} refers to the widely used monotonic neural networks, the min-max model \cite{danielsMonotonePartiallyMonotone2010a}, while \texttt{"Taylor"} represents an unconstrained Taylor NN model. Our proposed methods for monotonicity-informed neural networks include \texttt{"Mono Taylor"}, which incorporates monotonic constraints within the architecture NN, and \texttt{"Soft Taylor"}, which enforces monotonic constraints through the loss function. Additionally, \texttt{"1st"} and \texttt{"2nd"} represent the first- and second-order Taylor expansions. 

In the case of HVAC system modeling, the results presented in Table \ref{table} indicate that our monotonic Taylor NN models, including \texttt{"Mono Taylor"} and \texttt{"Soft Taylor"}, outperform the other models. Their $R^2$ scores consistently exceed 90\% across five multi-step prediction horizons, and their $RMSE$ values remain the smallest and most stable. Furthermore, the results demonstrate that higher-order Taylor approximations yield more accurate predictions. For instance, the performance of the \texttt{"Min-max"} model surpasses that of the \texttt{"1st Taylor"} model, yet falls short compared to the \texttt{"2nd Taylor"} model.  Moreover, the incorporation of physics-informed constraints improves accuracy even for lower-order models. The first-order version of our proposed models performs better than the second-order models without the physics-informed constraints. With the second-order model incorporating monotonicity constraints, the results are approximately similar to the first-order model; however, by the fifth prediction step, the second-order models exhibit superior results compared to the first-order models.

\subsection{Model predictive controller for TCLab}
\textbf{TCLab:}
The Temperature Control Laboratory \cite{rossiterEvaluationDemonstrationTake2019} is a system equipped with two heaters and two temperature sensors. Each heater is responsible for providing thermal energy to its corresponding sensor. This system can be classified as a multi-input, multi-output (MIMO) system, as it allows control over multiple inputs (heater powers) while monitoring multiple outputs (temperatures).
We denote the temperature measured by the first sensor as \(T_1\), with its corresponding heater power denoted as \(Q_1\). Similarly, the second sensor's temperature is denoted as \(T_2\), with its heater power represented as \(Q_2\). In addition to the affection of the heater to the sensor, the temperature of each sensor also interacts with each other. The TCLab system dynamics can be expressed through the following equation:
\begin{align}
    \begin{bmatrix}
        T_1^{k+1} \\[3pt]
        T_2^{k+1}
    \end{bmatrix}
    &= f_{\text{TC}}\left(
    \begin{bmatrix}
        (T_1^{k}, \hspace{0.15cm} Q_1^{k}, \hspace{0.15cm}T_2^{k}) \\[3pt]
        (T_2^{k}, \hspace{0.15cm}Q_2^{k}, \hspace{0.15cm}T_1^{k})   
    \end{bmatrix} \right). \label{TC_lab}
\end{align}

Based on the strong performance shown in Table \ref{table}, the \texttt{"1st Soft Taylor"} model was selected for implementation in the model predictive controller (MPC) for the real-world TCLab system. This model utilizes a first-order MTSNN, as described in \eqref{Taylor_NN}, and is trained using the loss function defined in \eqref{mon_mse}. During the data collection process for training, we randomly varied the heater values between \(10\%\) and \(50\%\) at intervals of 120 or 150 seconds. The resulting dataset, with a time step of 15 seconds, includes inputs \([T_1, Q_1, T_2, Q_2]\), corresponding to the temperatures and heater powers of both sensors. A total of \(N_{\text{train}} = 250\) data points were used for training.

\textbf{Testing scenario:}
The initial temperature of both sensors (\(Temp.\ 1\) and \(Temp.\ 2\)) is \(30^\circ C\). We set the reference temperature for sensor one (\(Ref.\ 1\)) to \(55^\circ C\), which creates a significant difference from the initial state. The reference temperature for sensor two (\(Ref.\ 2\)) is set to \(45^\circ C\).

The range of the training inputs for the \texttt{"1st Soft Taylor"} model is between \(10\%\) and \(50\%\). To test the model’s generalization ability, we design a model predictive controller with a maximum control input \(Q^{1,2}\) of \(65\%\), which is outside the training data range. Due to the high reference temperatures of sensors, we set the minimum control inputs to \(30\%\) for \(Q_1\) and \(20\%\) for \(Q_2\), respectively. 

\begin{figure}[t]
    \centering
    \includegraphics[width=0.95\linewidth, , trim=0.0cm 0.0cm 0.00cm 0.0cm, clip]{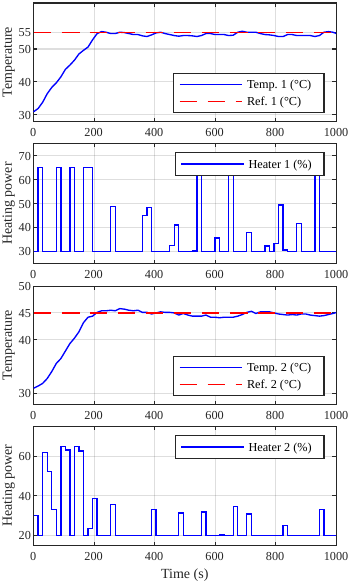}
    \caption{Measured temperature and control input in real TCLab using MPC}
    \label{fig:MPC_result}
\end{figure}

\textbf{Experimental results:}
\figref{fig:MPC_result} highlights that our control model efficiently drives the measured temperatures of each sensor toward their respective reference values ($55 ^\circ C$ for sensor one and $45 ^\circ C$ for sensor two) and maintains stability once these targets are reached, after 200 seconds (about 15 steps), despite the huge difference between the initial and reference temperatures ($25 ^\circ C$ and $15 ^\circ C$ for sensor one and sensor two).  On the other hand, when the temperature of two sensors adapts to their references, the heating power of heater two \(Q_2\) operates less than that of heater one \(Q_1\) due to the higher temperature of sensor one \(Temp.\ 1\) and temperature of sensor two \(Temp.\ 2\) being affected by \(Temp.\ 1\).
Moreover, the control inputs for both heaters frequently peak at around 
$65 \%$, which exceeds the range of the training data (from $10 \%$ to $50 \%$). Despite this, the measured temperatures remain stable, consistently hovering around their desired reference points. This demonstrates the robustness of our approach in generalizing beyond the training data range, a capability for practical control applications.

Consequently, the HVAC (Table \ref{table}) and TCLab system (\figref{fig:MPC_result}) results demonstrate that our proposed model performs accurately in system identification tasks, even when applied to experimental data. Furthermore, for the control aspect, the method exhibits robust performance in practical control applications, proving its generalization capability.

\section{Conclusion}
This study introduced a novel method for incorporating monotonicity or convexity constraints into machine learning algorithms using the Taylor series. These characteristics were integrated through both the neural network architecture and the loss function. Experimental results demonstrated that this approach outperformed unconstrained Taylor neural networks and the min-max model when tested with data from HVAC systems. Additionally, the proposed methods were successfully implemented in a real-world setting within the TCLab for designing model predictive controllers. Future work will aim to extend this approach to refine the distinction between subsequent and initial optimal points expansion in the Taylor series expansion. 


\end{document}